\documentclass[]{raa}            % referee version: for submission
\usepackage{graphicx,times}
\usepackage{amssymb,amsmath}
\usepackage{natbib}
\usepackage{enumerate}

\usepackage[breaklinks, colorlinks, urlcolor=red, filecolor=red, citecolor=blue, linkcolor=red, anchorcolor=red]{hyperref}

\newcommand{\SM}{\,{\ensuremath{M_{\odot}}}}
\newcommand{\SMPY}{\,{\ensuremath{M_{\odot}\,{\rm yr^{-1}}}}}

\def\DP#1{{\ensuremath{10^{\rm #1}}}}
\def\TDP#1{{\ensuremath{\times 10^{\rm #1}}}}

\begin{document}

\title{Unifying neutron star sub-populations in the supernova fallback accretion model}

\volnopage{Vol.0 (200x) No.0, 000--000}      %%preserved for Editor. DOn't remove!
\setcounter{page}{1}          %%starting page, preserved for Editor. DOn't remove!

\author{Bai-Sheng Liu
      \inst{1,2,3}
   \and Xiang-Dong Li
      \inst{1,2}
   }

\institute{Department of Astronomy, Nanjing University, Nanjing
210046, China; {\it lixd@nju.edu.cn}\\
%% Please give the E-mail address of the author, to whom future correspondence and
%% offprint requests will be sent.
        \and
Key Laboratory of Modern Astronomy and Astrophysics (Nanjing University), Ministry of Education, Nanjing 210046, China\\
        \and
Key Laboratory of Particle Astrophysics, Institute of High Energy Physics, Chinese Academy of Sciences, Beijing 100049, China\\
   }

   \date{Received~~2009 month day; accepted~~2009~~month day}

\abstract{We employ the supernova fallback disk model to simulate the spin evolution of isolated young neutron stars (NSs). We consider the submergence of the NS magnetic fields during the supercritical accretion stage and its succeeding reemergence. It is shown that the evolution of the spin periods and the magnetic fields in this model is able to account for the relatively weak magnetic fields of central compact objects and the measured braking indices of young pulsars. For a range of initial parameters, evolutionary links can be established among various kinds of NS sub-populations including magnetars, central compact objects and young pulsars. Thus, the diversity of young NSs could be unified in the framework of the supernova fallback accretion model.
}

\authorrunning{B.-S. Liu \& X.-D. Li }                % author_head in even pages
\titlerunning{Unifying neutron star sub-populations}  % title_head in odd pages

\maketitle

%________________________________________________ sections below
%
\section{Introduction}
\label{sec.1}

Neutron stars (NSs) are produced in core-collapse/electron capture supernovae (SNe) of massive stars or accretion-induced collapse (AIC) of massive white dwarfs. They were discovered as rotation-powered radio pulsars at first (\citealt{Hewish68,Pacini68,Gold68}). In pulsar astronomy, the spin evolution is one of the most prominent problems. Generally, the spin-down of isolated NSs can be described by a power-law (\citealt{ManchesterT77})
\begin{eqnarray}
  \dot{\Omega}=-K\Omega^{n},\label{eqn01}
\end{eqnarray}
where $\Omega$ and $\dot{\Omega}$ are the angular velocity and its time derivative, respectively, $K$ is a coefficient related to the spin-down torque, and $n$ is the braking index ($n=\Omega\ddot{\Omega}/\dot{\Omega}^2$ if $K$ is constant). In the case of pure dipole radiation, $K=2B^2R_{\rm NS}^6\sin^2\alpha/3Ic^3$ and $n=3$ (\citealt{ShapiroT83}). Here $B$, $R_{\rm NS}$, $\alpha$, and $I$ are the surface magnetic field strength, the radius, the angle between the magnetic and rotational axes, and the moment of inertia of the NS, respectively,  $c$ is the light speed. The characteristic age of a NS is then $\tau_{\rm c}=\frac{P}{(n-1)\dot{P}}[1-(\frac{P_0}{P})^{n-1}]$, where $P\,\, (\equiv 2\pi/\Omega)$ is the spin period, $\dot{P}$ is its time derivative and $P_0$ is the initial period.

Measurement of the braking indices was inhibited by timing noise, glitches, and limited spans of timing observations (e.g., \citealt{GavriilK04,DibKG08}), except for a handful of young pulsars with relatively stable long-term spin-down (\citealt{Espinoza13, GaoLiW16, Archibald16, GaoWSL17}, and references therein). Most of the measured braking indices $<3$, deviating from that in the magnetic dipole losses model. Many models have been developed to explain this discrepancy, involving the pulsar wind of high-speed particles causing loss of angular momentum from pulsars (\citealt{ManchesterT77, HardingE99, OuE16}), relativistic particles powered by a unipolar generator (\citealt{XuQ01}), variation in the dipolar inclination of the magnetic field (\citealt{ContopoulosS06, LyneE13}), the distribution of the nondipolar magnetic field (\citealt{BarsukovT10}), or the interaction between the NS and its fallback disk (\citealt{AlparAY01, MarsdenLR01, ChenL06, ChenL16, YanPS12, CaliskanE13, LiuE14}) which contributes a significant additional torque braking the pulsars besides dipolar radiation, a modified magnetodipole radiation model (\citealt{MagalhaesE12}), the growth of the magnetic field (\citealt{BlandfordR88, EspinozaE11, Ho15}), decrease of the effective moment of inertia as the superfluid core of the cooling NS increases (\citealt{HoA12}), etc. Especially two models are appealing to us, i.e., the field growth model and the fallback disk model, both of which have some interesting implications for the evolution of NSs and will be considered in this work.

Most pulsars possess magnetic fields in the range of $\sim 10^{11}-10^{13}$ G. Observationally there do exist NSs with ultra-strong fields ($\sim$ {\DP{14-15}} G) called magnetars (\citealt{Turolla09, Mereghetti11, OlausenK14, KaspiB17}) including anomaly X-ray pulsars (AXPs) and soft gamma-ray repeaters (SGRs). X-ray emission of these objects is thought to be powered by magnetic energy rather than rotational kinetic energy (e.g., \citealt{ThompsonD96}). Meanwhile some young NSs called central compact objects (CCOs, so-called anti-magnetars) associated with supernova remnants (SNRs) possess relatively weak magnetic fields ($\lesssim$ {\DP{10-11}} G, \citealt{GotthelfHA13a, GotthelfHA13b, Ho13, TorresPons16, DeLuca17}, and references therein). And the type I X-ray burster Circinus X-1, recently shown to be located in a SNR (\citealt{HeinzSF13}), also points to the occurrence of a low magnetic field in young NSs.

The observational diversity demonstrates the complicated formation and evolutionary processes of young NSs. However, the fact that the sum of the birthrates of various kinds of Galactic NS sub-populations exceeds the SN rate (\citealt{KeaneK08}) implies that there are likely evolutionary links among different NS species (e.g., \citealt{Kaspi10, PopovPP10}). A key process is the magneto-rotational evolution including either field decay (\citealt{Gullon14,Gullon15}) or growth (\citealt{Ho15}). However, these models usually focused on partial NS sub-populations (\citealt{Kaspi10, ViganoP12}). For example, by building up the association of the thermal evolution with magnetic field decay (i.e., the magnetothermal evolution), the field-decay model provided natural evolutionary tracks for NSs with high fields ($\gtrsim$ {\DP{13}} G, \citealt{Kaspi10, Gullon14, Gullon15}), but hardly explained the deviation of pulsar braking indices from that in the pure dipolar model. The field-growth model was suggested to account for the low fields of CCOs (e.g., \citealt{ViganoP12}), the evolutionary linkage between high-$B$ pulsars and magnetars (\citealt{EspinozaE11}), and the deviation of pulsar braking indices in young pulsars (e.g., \citealt{Ho15}). We note that there are few works to combine both processes in a unified picture. Here we assume that fallback accretion plays an important role in the magneto-rotational evolution of young NSs, and investigate the evolution of these objects in such a model.

It has been proposed that, during core-collapse SNe (\citealt{Colgate71}) or AICs (\citealt{DessartBY06}), part of the ejected matter may be bound to the newborn NSs and fall back. Then a disk  may form, if the specific angular momentum of the bound material is larger than the Keplerian value at the NS surface (\citealt{AlparAY01, MarsdenLR01, MenouPH01,EksiHN05}). The fallback disk systems, likely radiating in infrared (IR) or longer wavelengths during most of their lifetimes (e.g., \citealt{FosterF96, PernaHN00, PernaH00}), have been detected in a few cases (e.g., \citealt{IsraelE03, WangCK06, KaplanCWW09}).

If fallback disks are universal around young NSs, they may influence both the spin and magnetic field evolution, the degree of which depends on the amount of mass in the disks. The accretion/braking torque exerted by the fallback disk can spin up/down the NS with efficiencies much higher than that by magnetic dipole losses (e.g., \citealt{Chat00,ErtanEEA09}). \cite{YanPS12} and \cite{FuLi13} found that the torque exerted by the disk would greatly affect late-time period distribution depending on the initial parameter distribution. Especially, the extremely long spin period ($\sim 6.67$ hours, \citealt{DeLucaE06}) of 1E 161348$-$5055 (1E1613), a magnetar candidate (\citealt{ReaE16, DAiE16}. at the age of a few {\DP{3}} yr, \citealt{ClarkC76, NugentE84, CarterE97}) in the SNR RCW\,103 (\citealt{TuohyG80, GotthelfE97}), could be explained by the fallback disk model (\citealt{HoA17})\footnote{The time derivative of its $P$, the largest one among all known isolated NSs ($\dot{P}\leq$ {1.6\TDP{-9}}s\,s$^{-1}$, \citealt{EspositoE11}), denotes not only its strong field combining its long spin period, but also a more efficient braking process rather than dipolar radiation. In addition, 1E1613 should have been a magnetar with $2<P<12$ s unless it was spun down by a remnant disk timely (\citealt{HoA17}). Namely, the fallback disk can efficiently slow down 1E1613 to current very long ${P}$.}. In addition rapid fallback accretion may bury the magnetic fields and decrease their surface strengths (\citealt{Chevalier89, BernalPW13}), and the buried fields will gradually reemerge through Hall drift and Ohmic dissipation when accretion stops (see reviews by \citealt{Ho11, GourgouliatosC15}). Thus the coherent spin evolution will be substantially affected.

Based on \cite{LiuBSLi15}, here we build up a SN fallback disk model to study the NS spin evolution, where the field submergence and reemergence are both taken into account. Compared to previous studies, we have considered some important physical ingredients in the model. (i) The disk evolution satisfies a series of analytical self-similarity solutions (e.g., \citealt{ShenM12, LiuBSLi15}), depending on the disk status (e.g., slim disk or thin disk). Here self-gravity truncation and neutrality of the disk are both considered (see \citealt{LiuBSLi15}), which determines the age of the active disk and the influence of accretion on the NS spin. While other studies merely adopted a single power-law decline to describe the evolving disk (e.g., $\propto {t}^{-19/16}$ in \cite{YanPS12}, or a constant accretion rate in \cite{HoA17}), not to mention considering the self-gravity truncation or neutrality. (ii) In the early life of the NS, the magnetic field may undergo a quick burial. It is usually assumed that submergence of the magnetic fields lasts during the whole accretion phase (e.g. \citealt{FuLi13}). In our work, the field burial accords with the requirements as follows. Firstly, the magnetosphere should be compressed under the NS surface (\citealt{Muslimov95}). Secondly, both the total accreted mass and the accretion rate in the burial phase should exceed a critical value, respectively (\citealt{TorresPons16}).

The rest of the paper is organized as follows. Section \ref{sec.2} presents our theoretical considerations, followed by numerically calculated results in Section \ref{sec.3}. In Section \ref{sec.4}, we use the fallback disk model to describe the evolutionary links among various kinds of NS sub-populations. We summarize in Section \ref{sec.5}.

%%%%%%%%%%%%%%%%%%%%%%%%%%%%%%%%%%%%%%%%%%%%%%%%%%%%%%%%%%%%%%%%%%%%%%%%%%%%%%%%%%%%%%%%%%%%%%%%%%%%%%%%%%%%%%%%%%%
\section{The Model}
\label{sec.2}

Following the SN explosion, a SN shock is produced in the external layer of the star and propagates outwards. When the shock encounters the discontinuity in the outer H/He envelope, a reverse shock forms and travels inwards, driven by which some SN debris falls back (\citealt{Chevalier89})\footnote{The uncertainties of the fallback models can be summarized as follows (for a review, see \citealt{WongFE14}). Firstly, the mechanisms related to the fallback are still unclear. e.g., rarefaction wave deceleration and energy/momentum loss of the ejecta may be involved, besides the reverse shock deceleration. Secondly, the SN progenitors and the mass loss rates during the evolution of the progenitors are still under debate. In addition, different explosion-driven engines and numerical treatments used in the simulations can cause distinct amount of fallback.}. This is the supercritical accretion of the spherical fallback, and the fallback rate is highly super-Eddington (the typical rate is {\DP{-2}}$-${\DP{4}\SMPY}, \citealt{TorresPons16}).

According to studies on core-collapse SNe and AIC (e.g., \citealt{Fryer99, DessartBY06}), we assume that every newborn NS is surrounded by a fallback disk. Depending on its initial mass the newborn disk may be geometrically thick, and its mass-transfer rate be super-Eddington (\citealt{AbramowiczCLS88}). In the early phase the geometrical structure and form of the disk accretion are similar to those of the spherical accretion, respectively, even though the dynamics properties of the accretion flow in two scenarios are absolutely distinct. Therefore, we study the early disk accretion, and consider the influence of the field burial/reemergence and the disk evolution on the NS evolution.

Similar to \cite{LiuBSLi15}, the parameters were initiated as follows. (1) the NS mass $M=m {\SM}=1.4{\SM}$, and the radius $R_{\rm NS}=${\DP{6}} cm; (2) the mass flow rate $\dot{M}$ and the radii $R$ in the disk are scaled with the Eddington limit $\dot{M}_{\rm Edd}=1.39m$\TDP{18} gs$^{-1}$ and the Schwarzschild radius ${R}_{\rm S}=2.95m$\TDP{5} cm, respectively, i.e., $\dot{m}=\dot{M}/\dot{M}_{\rm Edd}$ and $r=R/R_{\rm S}$; (3) the timescale of the disk formation $t_{\rm f} \simeq$ 1.0$-$1000 ms, and the initial disk mass $M_{\rm D,0}=\eta{\SM} \simeq$ {\DP{-6}}$-$0.5{\SM}; (4) the initial inner and outer radii are set to be $r_{\rm i} \simeq$2.5 (i.e., $r_{\rm NS}=R_{\rm NS}/R_{\rm S}$) and $r_{\rm f} \simeq$10$-$1000, respectively. Hereafter the subscript 0 denotes parameters evaluated at $t_{\rm f}$. The newborn disk will expand owing to angular momentum transport, unless it becomes neutralized when its temperature is $<$ 1000 K (\citealt{KulebiE13}) or is truncated by self-gravity (See \citealt{LiuBSLi15} for the evolution of the outer radius, $r_{\rm out}$). The inner disk radius ($r_{\rm in}$) is determined by the interaction between the NS magnetic field and the disk.

\subsection{Spin Evolution}
\label{sec.2.1}

Based on \cite{LiuBSLi15}, the disk evolution can be divided into four phases. (i) In the 1st phase, the newborn disk evolves as a slim disk, where advective cooling is dominant and accretion rate is super-Eddington. (ii) In the 2nd phase, the outermost region gradually becomes geometrically thin and optically thick, where radiation is the main cooling source. (iii) In the 3rd phase, the entire disk becomes a thin disk. (iv) In the 4th phase, the inner region starts to turn into advection-dominated accretion flow (ADAF, where the primary cooling source is advection), unless the inner region is truncated by the NS magnetic field. Note that in phase 2 radiation pressure and electron scattering opacity in the outermost region are gradually replaced by gas pressure and free-free absorption, respectively. If the kinetic viscosity coefficient ($\nu$) is a power law of disk radius ($\nu \propto {R}^{n}$) solved from the energy and pressure equations for the accretion flow, the disk evolution in each phase should accord with the self-similarity solution (see e.g., \citealt{ShenM12, LiuBSLi15}). Thus the mass flow rate in the disk can be described as follows,
\begin{eqnarray}
   && \dot{m} = \left\{ \begin{array}{ll}
  \dot{m}_{0}(\frac{t}{t_{\rm f}})^{-4/3}  & (t_{\rm f}<t\leq{t}_{\rm 1}),   \\
  r_{\rm f}(\frac{t_1}{t_{\rm f}})^{38/21}(\frac{t}{t_{\rm f}})^{-8/7}          & (t_{\rm 1}<t\leq{t}_{\rm gas}), \\
  r_{\rm f}(\frac{t_1}{t_{\rm f}})^{38/21}(\frac{t_{\rm gas}}{t_{\rm f}})^{3/14}(\frac{t}{t_{\rm f}})^{-19/14}        & (t>t_{\rm gas}),
  \end{array}
  \right.\label{eqn02}
\end{eqnarray}
where $\dot{m}_0$ is determined by the initial disk mass (\citealt{CannizzoGeh09}), $t_1$ and $t_{\rm gas}$ are the times when the outer region of the disk starts to be dominated by radiative cooling and by gas pressure, respectively. Here the wind loss in the slim disk/ADAF is ignored for simplicity.

It is not precisely true that the NS accretion is only regulated by its radiation pressure. e.g., when the falling rate of the flow is larger than $\dot{M}_{\rm lim}=$ {3\TDP{-4}\SMPY} (or $\dot{m}_{\rm lim}=\dot{M}_{\rm lim}/\dot{M}_{\rm Edd} \simeq$ {\DP{4}}), photons will be trapped inside the flow and neutrino cooling be dominant (\citealt{Chevalier89}). The situation occurs in the newborn NS. Once $\dot{M}$ drops below $\dot{M}_{\rm lim}$, photons can diffuse outwards and the Eddington limit starts to work. Here we assume that $\dot{M}_{\rm lim}$ of the disk accretion equals that of the spherical accretion owing to their geometrical similarities (\citealt{Chevalier89, Chevalier96}). Hence the dimensionless accretion rate ($\dot{m}_{\rm NS}=\dot{M}_{\rm NS}/\dot{M}_{\rm Edd}$) of a NS with a fallback disk is given by,
\begin{eqnarray}
   && \dot{m}_{\rm NS} = \left\{ \begin{array}{ll}
  1.0       & (1.0 \leq \dot{m} < \dot{m}_{\rm lim} ),\\
  \dot{m}   & (\mbox{otherwise}).
  \end{array}
  \right.\label{eqn03}
\end{eqnarray}

The NS spin evolution depends on the magnitude of the magnetospheric radius $R_{\rm M}$, the corotation radius $R_{\rm co}$, and the light cylinder radius $R_{\rm LC}$. The magnetospheric radius is thought to be close to the Alfv\'{e}n radius for spherical accretion (\citealt{Ghosh07}, and references therein), at which the energy density of the accretion flow equals the magnetic energy density\footnote{Here $R_{\rm M}$ can be solved from the equilibrium equation of the magnetic energy and the kinetic energy of the accretion flow (\citealt{LambPP73}). In the supercritical accretion phase, although the most of the kinetic energy becomes thermal energy, the total energy density around $R_{\rm M}$ is hardly changed. And the radiation pressure can hardly affect the equilibrium of the pressure of the infalling flow with that of the magnetic field around $R_{\rm M}$, since the neutrino emission is dominant in the phase. Alternatively, $R_{\rm M}$ can also be obtained by equating the angular-momentum-transfer rate of the disk with the magnetic torque (\citealt{GhoshL79}), which is marginally different from Eq.~(\ref{eqn04}). If $\dot{M}$ is supercritical, the disk matter rotates at a sub-Keplerian angular velocity, i.e., $\Omega=A\Omega_{\rm K}(r)$, where $\Omega_{\rm K}$ is the Keplerian rate and $A<1$ (\citealt{NarayanY95}). Thus the magnetospheric radius is only varied by a factor of $A^{-2/7}$ (e.g., \citealt{AnderssonGHW05, XuLi17}), denoting a slight change of $R_{\rm M}$ if typical values of $A$ ($\sim 0.2-0.3$, \citealt{NarayanY95}) are adopted. In ADAF with low $\dot{M}$, we similarly ignore the factor of $A^{-2/7}$.}. The corotation radius is obtained by equating the angular velocity of the NS to the local Keplerian one. At the light cylinder radius the corotating velocity equals the light speed. These radii can be written as follows (\citealt{FrankKR02})
\begin{eqnarray}
  && r_{\rm M}=\frac{R_{\rm M}}{R_{\rm S}} \simeq 2.19 {m}^{-10/7} B_{8}^{4/7} \dot{m}^{-2/7}, \label{eqn04}\\
  && r_{\rm co}=\frac{R_{\rm co}}{R_{\rm S}} \simeq 17.32 (m\Omega)^{-2/3}, \label{eqn05}\\
  && r_{\rm LC}=\frac{R_{\rm LC}}{R_{\rm S}} \simeq 101.70 \Omega^{-1},\label{eqn06}
\end{eqnarray}
where $B_8$ is the NS magnetic field in units of $10^8$ G. Note that $r_{\rm M}<r_{\rm i}$ at high $\dot{m}$, thus the disk inner radius should be $r_{\rm in}=\max(r_{\rm M},\,\, r_{\rm i})$.

With the decline of $\dot{m}$, the NS may experience the following evolutionary stages.

(i) If $r_{\rm in}<r_{\rm co}$, the disk material can fall onto the star surface. The NS spin-up/down depends on the angular momentum of the accretion flow and its interaction with the magnetic field (\citealt{FrankKR02})
\begin{equation}
  I\dot{\Omega}=\dot{M}_{\rm NS}(GMR_{\rm in})^{1/2}(1-\frac{\omega_{\rm s}}{\omega_{\rm c}}),\label{eqn07}
\end{equation}
where $\omega_{\rm s}=\Omega/\Omega_{\rm K}(R_{\rm in})=(R_{\rm in}/R_{\rm co})^{3/2}$ is the fastness parameter, $\Omega_{\rm K}(R_{\rm in})$ is the Keplerian angular velocity at $R_{\rm in}$, and $\omega_{\rm c}=0.5-1.0$ (\citealt{Ghosh07}). If $\Omega=\Omega_{\rm K}(R_{\rm in})$, the NS will spin at the equilibrium period, reading $P_{\rm eq}=2\pi/\Omega_{\rm K}(R_{\rm in})$. In the following calculation, $\omega_{\rm c}=0.7$ is assumed.

(ii) If $r_{\rm co}<r_{\rm in}<r_{\rm LC}$,  the system is in the propeller state, and the disk matter can hardly be accreted by the NS. The spin evolution of the NS is also described by Eq.~(\ref{eqn07}). Since $\omega_{\rm s}>1$ in the state, the propeller effect should cause the star to spin down (\citealt{dan2010}). Especially, when $\omega_{\rm s}\gg 1$, the spin-down torque in Eq.~(\ref{eqn07}) will recover to that of \cite{ill1975}.

(iii) When $\dot{m}$ decreases so that $r_{\rm in}>r_{\rm LC}$, i.e., the disk is expelled outside of the light cylinder and the NS is in the ejector phase. It is assumed that radio activity operates in NSs with normal fields and those with super-strong fields appear as magnetars in this situation. In either case the spin-down is caused by the magnetic dipole losses.

\subsection{Magnetic Field Evolution}
\label{sec.2.2}

The submergence process of the magnetic field has been investigated in detail (\citealt{TorresPons16}), which has been used to account for such observations as the remarkably low fields of CCOs (\citealt{GotthelfHA13a, GotthelfHA13b}), and frequent thermonuclear bursts from the X-ray binary Circinus X-1 (\citealt{HeinzSF13}). The field screening, with a duration of a few hours (\citealt{TorresPons16}), happens in the supercritical accretion phase (\citealt{Chevalier89, Muslimov95}).

During the supercritical accretion phase, the accretion fluid interacts with the magnetosphere. Before reaching the magnetosphere, the accretion flow falls supersonically, and hence an accretion shock forms inevitably above the magnetopause, which diffuses outwards. And the shocked flow would fall subsonically at the base of the accretion shock. Finally the magnetopause may reach a lower equilibrium radius and be screened by the accreted flow. Here we utilize the field burial caused by spherical accretion to study that by disk accretion in the early phase, since they have similar properties close to the surface of the NS (\citealt{Chevalier96}).

The occurrence of field screening depends on the factors as follows. First of all, for a successful field-submergence the magnetosphere should be compressed beneath the NS surface (i.e., $R_{\rm M} \leq R_{\rm NS}$ and $R_{\rm M} \leq R_{\rm co}$, \citealt{Muslimov95}). Secondly, during the field decay, the accretion rate $\dot{M}$ should surpass a critical burial rate ($\dot{M}_{\rm CB}$), and the effective mass ($\Delta{M}$) accreted to bury the NS magnetic field exceed a critical value ($M_{\rm cr}$) (\citealt{TorresPons16}). That is, during the accretion phase, only when $R_{\rm M} \leq R_{\rm NS}$ and $\dot{M} > \dot{M}_{\rm CB}$, can the accreted mass contribute to the field burial and add to $\Delta{M}$. Moreover, the field submergence will never begin unless $\Delta{M} > M_{\rm cr}$. Here the critical mass is approximated as
\begin{eqnarray}
  \lg\frac{{M}_{\rm cr}}{\SM}=1.5(\lg\frac{B_{0}}{\rm {\DP{14}}\,G}+0.48\lg\frac{t_{\rm ac}}{\rm 1\,yr})+2.61, \label{eqn08}
\end{eqnarray}
which depends on the initial magnetic field $B_{0}$ and the effective accretion timescale ($t_{\rm ac}$, approximately estimated from $R_{\rm M} = R_{\rm NS} \leq R_{\rm co}$)
\begin{eqnarray}
   && \lg({t}_{\rm ac}+{t}_{\rm f}) = \left\{ \begin{array}{ll}
  \lg{t}_{\rm f}+\frac{3}{4}(\lg\dot{m}_{0}- 3.5\lg{f_{\rm P}})  & (t_{\rm f}<t\leq{t}_{\rm 1}),   \\
  \lg{t}_{\rm f}+\frac{7}{8}(\lg{k}_{2}- 3.5\lg{f_{\rm P}})      & (t_{\rm 1}<t\leq{t}_{\rm gas}), \\
  \lg{t}_{\rm f}+\frac{14}{19}(\lg{k}_{5}- 3.5\lg{f_{\rm P}})    & (t>t_{\rm gas}),
  \end{array}
  \right.\label{eqn09}
\end{eqnarray}
where $f_{\rm P}\approx 0.65{m}^{-3/7}(B_{0}/{\DP{8}}\,{\rm G})^{4/7}$, $k_{2}=r_{\rm f}(t_{1}/t_{\rm f})^{38/21}$, and $k_{5}=k_{\rm 2}(t_{\rm gas}/t_{\rm f})^{3/14}$ (see \citealt{LiuBSLi15}). And $\dot{M}_{\rm CB}={M}_{\rm cr}/{t}_{\rm ac}$ (see Fig. 7 in \citealt{TorresPons16}).

Obviously, the situation here is distinct from that in \citet{FuLi13}, where the field burial proceeds all the way if the NS is in the accretor phase. Here we refer the details of the burial process to \cite{ViganoP12} and \cite{TorresPons16}, and utilize an empirical relation as follows for field decay (\citealt{TaamHeuval86, ShibazakiMSN89, Romani90})
\begin{equation}
 B=\frac{B_0}{1+\Delta{M}/{\DP{-5}\SM}},\label{eqn10}
\end{equation}
where $\Delta{M}$ is the effective mass accreted during the burial stage. We also assume that the magnetic field will keep constant if no burial develops in the accretion phase, e.g., i) $R_{\rm M} > R_{\rm NS}$; ii) $\dot{M} < \dot{M}_{\rm CB}$; iii) $\Delta{M} < M_{\rm cr}$.

When accretion terminates, the screened field will slowly propagate outwards to the NS surface via Ohmic diffusion and Hall drift (\citealt{GeppertE99, Ho11, PonsVR13, ViganoRM13, GourgouliatosC15}). And the typical timescale of the reemergence is $\gtrsim$ {\DP{3}} yr (e.g., \citealt{Ho11, GourgouliatosC15}), depending on the burial depth. Here we adopt the following relation, fitted from the numerical results by \cite{ViganoP12}, to describe the field diffusion,
\begin{eqnarray}
  && {B(t)}={B}_{\min}\frac{k+1}{k \cdot{\exp}{[(t-t_{\rm G})/\tau_{\rm B}]}+1}+{B}_{0}(1-\frac{k+1}{k \cdot{\exp}{[(t-t_{\rm G})/\tau_{\rm B}]}+1}). \label{eqn11}
\end{eqnarray}
where ${B}_{\min}$ is the minimum field during the field decay, $\tau_{\rm B}\sim${\DP{2}}-{\DP{3}} yr (here we set $\tau_{\rm B}=$300 yr), $\lg{k}=$ $-[\lg\frac{{B}_{\rm 0}}{{B}_{\min}}]-1$, and $t_{\rm G}$ denotes time when the field growth begins. Here Eq.~(\ref{eqn11}) comprises two components, the decay of the surface field and the diffuse of the buried field, which accords with \citet{ViganoP12}. Since ${B}_{\min}$ and ${k}$ are both decreasing functions of $\Delta{M}$ (see Eq.~(\ref{eqn10})), accordingly with the increase of the effective accretion mass the field diffusion (the main component in Eq.~(\ref{eqn11})) becomes slower and slower.

Note that Eq.~(\ref{eqn11}) merely describes the field evolution within {\DP{5}} yr (see Figure 6 in \citealt{ViganoP12})\footnote{ As for a longterm evolution, we recommend one to multiply the second term on the right-hand side of Eq.~(\ref{eqn11}) by a factor similar to $\frac{k+1}{k \cdot{\exp}{[(t-t_{\rm G})/\tau_{\rm B}]}+1}$, which is an order of unity and hence ignored in our simulations.}, we terminate calculations at the age {\DP{5}} yr for NSs with $B_{0} \lesssim$ {\DP{14}} G and {\DP{4}} yr for those with $B_{0} \gtrsim$ {\DP{14}} G, respectively (\citealt{Ho15}). While within the timescale it is adequate for us to illustrate the early evolution of NSs, e.g., the weak magnetic fields of CCOs, the braking indices $<$ 3, and evolutionary links among NS sub-populations.

%%%%%%%%%%%%%%%%%%%%%%%%%%%%%%%%%%%%%%%%%%%%%%%%%%%%%%%%%%%%%%%%%%%%%%%%%%%%%%%%%%%%%%%%%%%%%%%%%%%%%%%%%%%%%%%%%%%
\section{Example Results}
\label{sec.3}

We construct a set of models based on the following parameters: the initial mass ($\eta${\SM}), the formation timescale ($t_{\rm f}$), the standard \citet{ShakuraS73} viscosity parameter ($\alpha$, equal to 0.1 here), the initial inner radius ($r_{\rm i}$) and the initial outer radius ($r_{\rm f}$) of the disk, the initial magnetic field strength ($B_{0}$) and the initial spin period ($P_{0}$) of the NSs. In order to simplify the description, we label each model with $\eta$, $t_{\rm f}$, $r_{\rm i}$ and $r_{\rm f}$, where $r_{\rm i}$ ($r_{\rm f}$) is absent unless $r_{\rm i}\ne$2.5 ($r_{\rm f}\ne$1000). For example, $\eta$0.3$t_{\rm f}$50 denotes a model with $\eta=0.3$, $t_{\rm f}= 50$ ms, $r_{\rm i}=2.5$ and $r_{\rm f}=1000$.

\begin{figure}%[ht]
\begin{center}
  \includegraphics[width=0.8\textwidth]{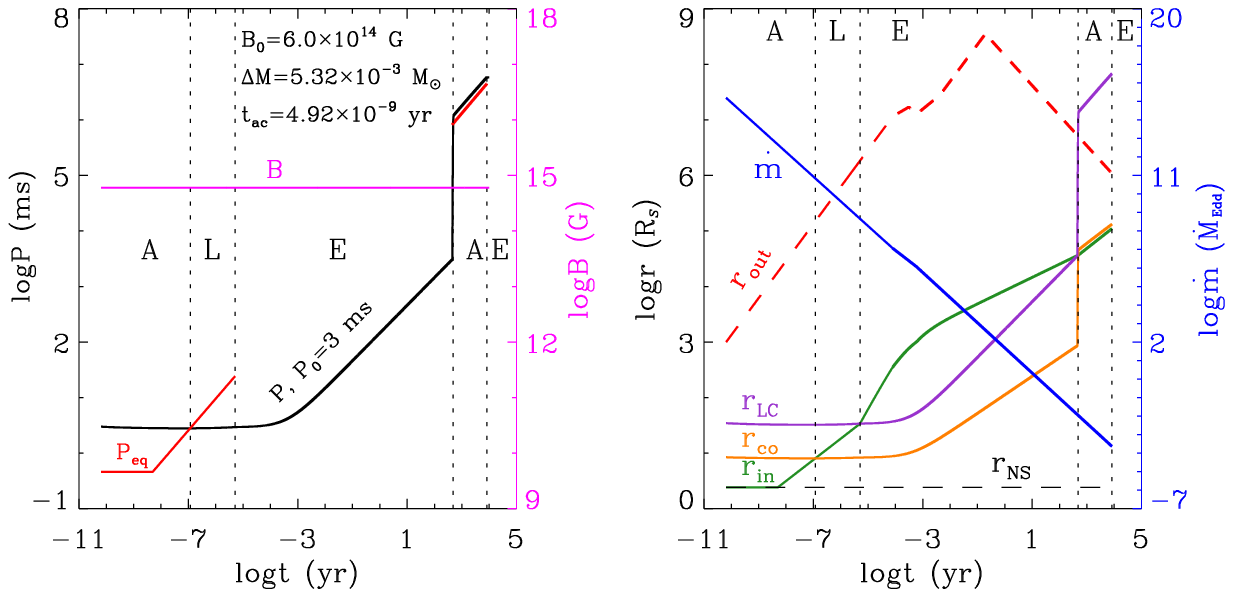}
  \caption{Evolution of the NS spin period (left panel) and several characteristic radii (right) in model $\eta${1.6\TDP{-2}}$t_{\rm f}$2. Some initial parameters are depicted in the left panel. Also shown are the magnetic field (left), the equilibrium period ($P_{\rm eq}$, left) and accretion rate ($\dot{m}$, right) during different phases. Here we separate these phases using black vertical dotted lines, including the accretor phase (labeled by A), the propeller phase (L) and the ejector phase (E).}
  \label{fig01}
\end{center}
\end{figure}

We firstly consider model $\eta${1.6\TDP{-2}}$t_{\rm f}$2. Fig.~\ref{fig01} shows the evolution of the spin period and the equilibrium period $P_{\rm eq}$ (left panel) and several important radii (right) for a magnetar-like NS with $B_0=6\times 10^{14}$ G and $P_0=$ 3 ms. In the figure, black vertical dotted lines divide different evolutionary phases, marked by A (the accretor phase), L (the propeller phase) and E (the ejector phase). At very high accretion rate ($\dot{M} \gtrsim${\DP{12}} $\dot{M}_{\rm Edd}$, see right panel), the magnetosphere is smaller than the star's radius, but the accreted effective mass ($\Delta{M}=${5.3\TDP{-3}\SM}, at accretion rate $\dot{M} \geq \dot{M}_{\rm CB}$) is less than the critical mass ($M_{\rm cr}=${6.3\TDP{-3}\SM}, within the effective accretion timescale $t_{\rm ac}=${4.9\TDP{-9}} yr), thus no field burial can occur in this case. As shown in the figure, the NS quickly switches from an accretor to a propeller, and then an ejector, based on the relationship of $r_{\rm in}$, $r_{\rm LC}$ and $r_{\rm co}$ (right panel). At the age $\sim 500$ yr an equilibrium period is reached, and the NS resides in the tracking phase with $P\simeq P_{\rm eq}$, observed as an X-ray source. In the right panel the outer radius $r_{\rm out}$, measuring the outer edge of the viscous disk ($\geq$ 1000 K), is plotted. Since the age $\sim$ 1 month, neutralization ($\leq$ 1000 K, \citealt{KulebiE13}) starts from the outer edge of the disk and develops inwards. Additionally, $r_{\rm in}$ increases with decreasing $\dot{m}$, which further truncates the inner hot region. We assume that the star will behave as a radio pulsar owing to the cease of accretion at age $\lesssim$ {\DP{4}} yr when $r_{\rm out}<10r_{\rm in}$ (\citealt{KulebiE13}).

\begin{figure}%[ht]
\begin{center}
  \includegraphics[width=0.7\textwidth]{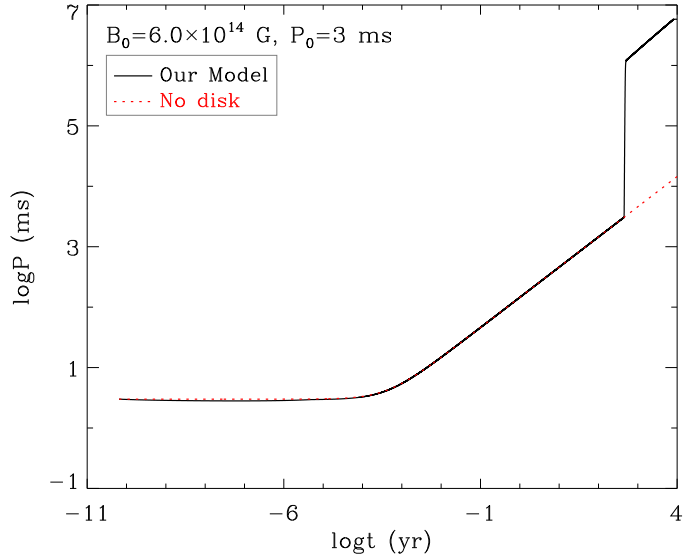}
  \caption{Same parameters as those in Fig.~\ref{fig01}. Here NS spin evolution without disk-assisted torque (red dotted line, i.e., no disk forms) is compared with our model.}
  \label{fig02}
\end{center}
\end{figure}

Fig.~\ref{fig02} compares the spin evolution with and without a fallback disk. Here no disk denotes no fallback happens. If no fallback disk forms, the spin evolution will be distinct from that in Fig.~\ref{fig01}, especially in the late phase. Firstly, the NS will never be braked by the disk effectively, and hence slow down at a much lower $\dot{P}$ (see evolution at $t\gtrsim$ 500 yr), finally reaching a much smaller spin period of $\sim$ 10 s ($<$ several hours in the fallback disk model). Secondly, the NS will be observed as a radio pulsar, rather than an X-ray source (caused by the material accretion). These distinctions are determined by the intensity of the interaction between the accretion flow and magnetic field.

\begin{figure}%[ht]
\begin{center}
  \includegraphics[width=0.8\textwidth]{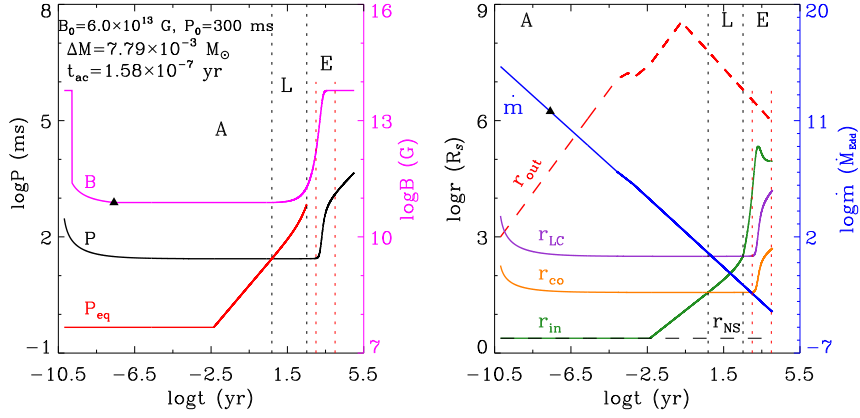}
  \caption{Similar calculation in model $\eta${1.6\TDP{-2}}$t_{\rm f}$2 to that in Fig.~\ref{fig01}. Here the NS undergoes a field burial, of which the endpoint is denoted by a black filled triangle. And two red vertical doted lines denote $t=${\DP{3}} and {\DP{4}} yr, respectively.}
  \label{fig03}
\end{center}
\end{figure}

We demonstrate the evolution of a lower-field NS in Fig.~\ref{fig03} based on model $\eta${1.6\TDP{-2}}$t_{\rm f}$2. The other parameters are $P_0=300$ ms and $B_0=${6\TDP{13}} G. The reasons why the field burial happens are illustrated as follows. (1) During the supercritical accretion ($t\lesssim$ {\DP{-2.5}} yr, see right panel), $R_{\rm M} \leq {R}_{\rm NS}$. (2) At $\dot{M} \geq \dot{M}_{\rm CB}$, $\Delta{M}=${7.8\TDP{-3}\SM} (slightly smaller than total mass accreted during accretion phase, $\sim$ {9.0\TDP{-3}\SM}), which outmatches ${M}_{\rm cr}=${2.4\TDP{-3}\SM}. Hence the surface field is screened down to {7.7\TDP{10}} G, which terminates once $\dot{M} < \dot{M}_{\rm CB}$ (indicated by the black filled triangle in the figure). With the increase of $R_{\rm in}$ ($\propto \dot{M}^{-2/7}$), the NS passes through the propeller phase to enter the ejector phase from the age $\sim$ 300 yr. Thereafter, accretion (halting at age $\lesssim$ {\DP{4}} yr) actually has no effect on its subsequent evolution since $r_{\rm LC}<r_{\rm M}$. It takes $\sim$ 3000 yr for the NS to recover to its initial field. This case may illustrate the formation of some high-$B$ radio pulsars.

\begin{figure}%[ht]
\begin{center}
  \includegraphics[width=0.8\textwidth]{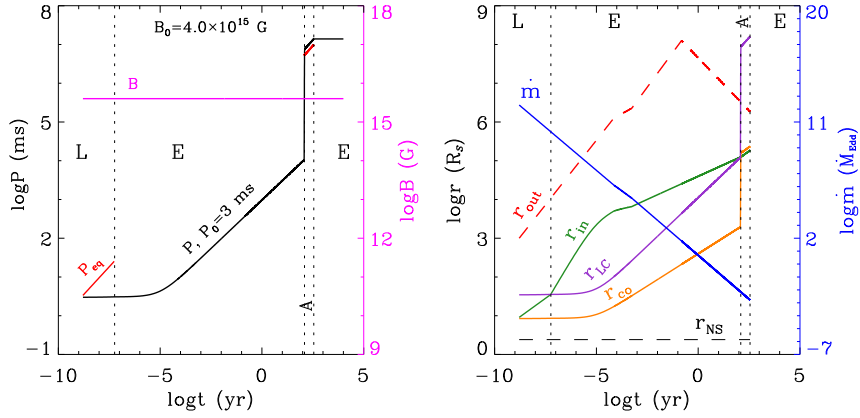}
  \caption{Evolution of a NS with a very strong field in model $\eta${5.3\TDP{-4}}$t_{\rm f}$50.}
  \label{fig04}
\end{center}
\end{figure}

In Fig.~\ref{fig04} we change the parameters to be $B_0=${4\TDP{15}} G and $P_0=3$ ms in model $\eta${5.3\TDP{-4}}$t_{\rm f}$50. Here the very strong field can hardly be buried since $R_{\rm M} > R_{\rm NS}$ in the early evolution. In the figure, the NS evolves from a propeller to an ejector, and subsequently approaches the tracking phase at age $\gtrsim$ 100 yr. About 300 yr later, the NS becomes a radio pulsar since the accretion turns off. Compared with Fig.~\ref{fig01}, due to the stronger field, the NS in this case can more effectively reach the spin period of several hours within $\sim$ 100 yr. Thus the model might explain the very long spin period ($\sim$ 6.67 hr) of 1E 1613 (\citealt{Li07, PizzolatoE08, BhadkamkarG09, IkhsanovKB15, HoA17, TendulkarE17}), a young magnetar candidate (with age of $\sim$ a few {\DP{3}} yr, see, e.g., \citealt{GotthelfE97}).

\begin{figure}%[ht]
\begin{center}
  \includegraphics[width=0.8\textwidth]{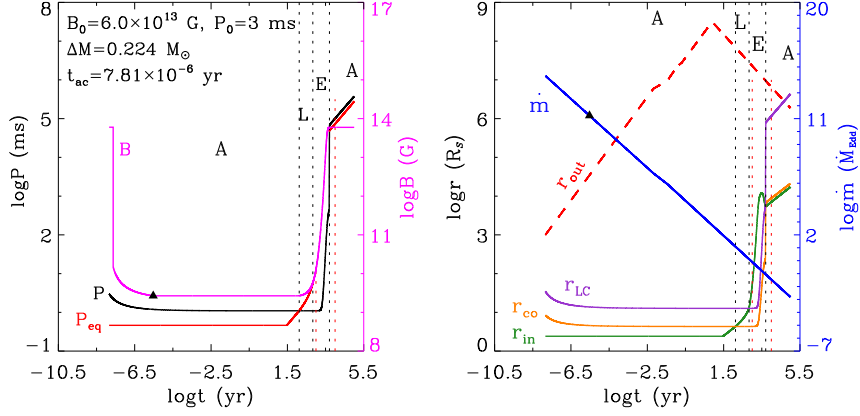}
  \caption{Evolution of a NS, undergoing a very heavy field burial, in model $\eta${0.48}$t_{\rm f}$450.}
  \label{fig05}
\end{center}
\end{figure}

A more extreme example is a simulation based on model $\eta${0.48}$t_{\rm f}$450 (see Fig.~\ref{fig05}). Here the magnetic field is deeply buried by the accretion flow with an effective mass $\sim$ 0.224{\SM}, with its magnitude decreasing from $6\times 10^{13}$ G to a few $10^{9}$ G. The NS evolves from an accretor to a propeller, and from an ejector to an accretor finally. Before the recovery of the buried field, the torque from accretion flow or magnetic dipole losses is too trivial to vary the NS rotation significantly, and the NS may behave as a CCO. With the retrieval of the buried field, the NS reaches the spin equilibrium at the age $\sim$ 5000 yr, observed as an X-ray pulsar. And the NS may spend long time at the tracking phase unless the accretion expires.

Finally, we show how the disk mass ($\eta${\SM}) or its formation timescale ($t_{\rm f}$) affects the field burial in Fig.~\ref{fig06}. In the $\eta-t_{\rm f}$ parameter space, the required field decay for NSs with different $B_{0}$ and distinct extent of field decay for those with specified $B_{0}$ are illustrated. The magenta line stands for $B_{0}=$ {\DP{14}} G, blue for {4\TDP{13}} G and black for {4\TDP{12}} G. The solid, dotted and dashed lines denote $\frac{B_{\min}}{B_0}=$ {\DP{-2}}, {\DP{-3}} and {\DP{-4}}, respectively, where $B_{\min}$ is the magnetic field at the end of the burial. According to our calculations, (1) NSs assuming a higher $B_0$ tend to undergo a more significant weakest-decay, especially when $t_{\rm f}$ exceeds some limit (see Eq.~(\ref{eqn08})--(\ref{eqn10})). e.g., for the burial of field $B_0=$ {\DP{14}} G, it requires that $\max(\frac{B_{\min}}{B_0})<${\DP{-3}} at $t_{\rm f} \gtrsim$ 240 ms or $\max(\frac{B_{\min}}{B_0})<${\DP{-2}} at $t_{\rm f} \gtrsim$ 2 ms. (2) The value of $\eta$ is nearly constant for specified $B_{\min}$ and $B_0$. Obviously significant field decay requires extensive supercritical accretion. If future numerical simulations of SN explosions reveal plausible distribution of the fallback parameters, one can estimate to what extent the NS fields can be influenced by the SN explosions with different progenitor masses.

\section{A Unified Picture}
\label{sec.4}

Figure~\ref{fig07} depicts the $B-P$ diagram for normal pulsars (dots), pulsars with measured braking indices (crosses), magnetars (stars), and CCOs (asterisks)\footnote{Data taken from \url{http://www.atnf.csiro.au/people/pulsar/psrcat/}, see also \cite{ManchesterHE05}.}. Based on the fallback disk model and numerical calculations presented in the former section, we demonstrate six representative evolutionary tracks (the related parameters are listed in Table~\ref{tab01}), trying to account for various species of young NSs ($\lesssim$ {\DP{6}} yr) in the framework of one unified picture. Our main point is that all newborn NSs are subject to the fallback accretion, which influences both the evolution of the magnetic field and the spin. The extent of that influence depends on the initial parameters of the fallback disk and the NS. Note that these evolutionary routes are illustrative rather than exact simulations for specific objects, starting from the formation of the disk to \DP{5} yr and {\DP{4}} yr for NSs with $B_{0} \lesssim$ {\DP{14}} G and $\gtrsim$ {\DP{14}} G, respectively. Generally NSs at first experience rapid accretion and field burial (shown with the solid lines), and then evolve into the propeller/ejector phase (the dotted/dashed lines).

Models M01, M02 and M03 describe the possible evolution of magnetars and high-$B$ pulsars. In model M01 the fallback accretion hardly suppresses the magnetic field, and the star spin is mainly braked by magnetic dipolar radiation. Thus model M01 may stand for the evolution of magnetars born with a disk $\lesssim$ {\DP{-2}\SM} (see Fig.~\ref{fig06}). In models M02 and M03, the field is firstly screened by supercritical accretion and then gradually grows to its original level. During the process, the NS may initially behave as a high-$B$ pulsar, and finally evolve into a magnetar. Hence these two models represent a possible evolutionary link between high-$B$ pulsars and magnetars. Additionally, unlike model M02, model M03 may demonstrate the formation of magnetars with a little lower magnetic field ($<$ {\DP{14}} G). Models M04 and M05 illustrate the evolutionary tracks of normal pulsars. Here model M05 may be representative for most radio pulsars. That is, small mass is accreted causing a weak field decay, and the dipole radiation torque is dominant during the star spin-down. In model M06 the field is deeply buried by a large amount of accretion mass, and thus during the slow growth of the field NSs may behave as CCOs. Moreover, for those in which the reemergence of the buried field has not finished, the braking indices are expected to be less than 3, as depicted in the figure (the filled triangle). In model M03 or M04, since the field growth mainly appears in the ejector phase, the NS should not be braked by the disk. Namely, here the braking index $<3$ is completely caused by the field growth, i.e., the second term on the right-hand side of Eq.~(\ref{eqn11}) since its time derivative is the main component in the evolution.

Magnetic fields of observed young NSs range from $\lesssim$ {\DP{10-11}} G (e.g., CCOs, \citealt{GotthelfHA13a, GotthelfHA13b}) to $\sim$ {\DP{14-15}} G (magnetars, e.g., \citealt{OlausenK14}), while in our unified model the initial fields accord with a usual distribution, i.e., $\gtrsim$ {\DP{12}} G. By taking into account the fallback accretion, our model could not only describe the birth and evolution of various NS sub-groups, but also illustrate their evolutionary link and the braking indices $< 3$.

%%%%%%%%%%%%%%%%%%%%%%%%%%%%%%%%%%%%%%%%%%%%%%%%%%%%%%%%%%%%%%%%%%%%%%%%%%%%%%%%%%%%%%%%%%%%%%%%%%%%%%%%%%%%%%%%%%%
\section{Discussion and Conclusions}
\label{sec.5}

Young NSs can manifest themselves as magnetars, high-$B$ pulsars, CCOs, etc. However, the classification of these NSs becomes ambiguous in some cases. The high-$B$ pulsars could be the quiescent counterparts of the transient magnetars deduced from their similar radio spectra (\citealt{KaspiMc05, CamiloE06, CamiloE07}), two high-$B$ pulsars (PSR J1846$-$0258, \citealt{GavriilE08}; PSR J1119$-$6127, \citealt{ArchibaldE16, GogusE16}) and a CCO candidate (1E 161348$-$5055, \citealt{TuohyG80, GotthelfE97}) displayed magnetar-like X-ray bursts (\citealt{ReaE16, DAiE16}). Moreover, there must be evolutionary links among at least some NS sub-populations in order to account for the birthrate problem (\citealt{KeaneK08}). There have been studies trying to figure out a unified model to describe observations of these populations. In terms of the magnetothermal evolution model (\citealt{PopovPP10, PonsVR13, Gullon14}), some of NS populations could be connected with similar evolutionary paths but with different initial field configurations (see \citealt{Kaspi10}, for a review). Especially, based on current observations of radio pulsars and thermal X-ray emission of X-ray pulsars, the magnetothermal evolution (\citealt{ViganoRM13}) and modified magnetospheric models are combined to reproduce the birthrate and birth properties of NS population (e.g., the distribution of the spin and the magnetic field, \citealt{Gullon15}).

Even so, two issues remain to be resolved. Firstly, the formation of CCOs, which are slightly younger than AXPs/SGRs but with much weaker field strengths (e.g., \citealt{GotthelfHA13a, DeLuca17}), is difficult to understand. If all objects should undergo a magnetic field amplification during the proto-NS phase (caused by the rapid rotation, convection and/or turbulence, see discussion in \citealt{TorresPons16}), CCOs should have accreted much more material to submerge its amplified fields than other isolated NSs. Secondly, in the framework of magnetic dipole losses, the measured braking indices are mostly less than 3 (\citealt{Espinoza13}. except that of PSR J1640$-$4631, \citealt{Archibald16, GaoWSL17}), which is not readily accounted for by the magneto-thermal evolution model. There are several plausible explanations for the deviation of measured braking indices, among which the reemergence of the buried field can supply a promising solution (e.g., \citealt{BlandfordR88, GourgouliatosC15}).

Obviously, none of the above models described all NS families in a same framework. The magnetothermal evolutionary model is mainly applicable for NSs with high magnetic fields ($\gtrsim$ {\DP{13}} G, e.g., AXPs/SGRs, high-$B$ pulsars. See \citealt{PonsE07, PopovPP10, Kaspi10, PonsVR13, ViganoRM13, Gullon14, Gullon15}), the decay of which produces significant heat and hence creates a connection with thermal evolution. Some other studies tried to unify NSs with low and high surface-fields in the field-burial model, believing that CCOs are `hidden' magnetars (e.g., CXO J1852.6+0040, 1E161348$-$5055. See \citealt{DeLuca08, PopovE15}). Especially, although the model in \cite{PopovE15} is similar to ours, the NSs involved in their work should assume very strong crustal fields, which may exclude partial CCOs. In addition, the field growth model was developed to illustrate very low fields of the CCOs (\citealt{ViganoP12}), the possibility of high-$B$ pulsars becoming magnetars (\citealt{EspinozaE11}), and the braking indices $<$ 3 of young pulsars (\citealt{Ho15}). If the field burial and growth are considered simultaneously, not only can the above two problems find a way out, but also the evolutionary links among CCOs, high-$B$ pulsars, and magnetars can be established (\citealt{RogersS16}).

No physical models has been constructed currently linking the field burial and growth, or unify all NS sub-populations. In our work we tentatively adopt the fallback accretion to drive the variation of the magnetic field, and related spin evolution of young NSs, taking into account the field burial/diffusion. Moreover, we introduce some new factors. Firstly, for each disk status (e.g., slim disk, thin disk), a self-consistent solution is adopted (see Eq.~(\ref{eqn02}), rather than a single one, $t^{-5/3}$). Here self-gravity truncation and neutralization of disk are both considered, which affect the disk diffusion, and determine whether the accretion is active. Otherwise, the disk outer region would diffuse freely, and the mass transfer in the disk would never turn off, unless the disk is destroyed by the magnetic field. Secondly, more stringent requirements of the field burial are introduced here according to state-of-art numerical simulations. i.e., not only should the magnetosphere be lower than the NS surface during the supercritical accretion (\citealt{Muslimov95}), but also the accretion mass and rate should both exceed a critical value, respectively (\citealt{TorresPons16}). Compared with \cite{FuLi13}, the NS should undergo a weaker or even no field decay in our model. Lastly, here a universal equation, fitted from simulations in \cite{ViganoP12}, describes the growth of the screened field. Based on our simulations, possible evolutionary tracks connecting various types of NSs are found. We summarize the results as follows.

\begin{figure}%[ht]
\begin{center}
  \includegraphics[width=0.7\textwidth]{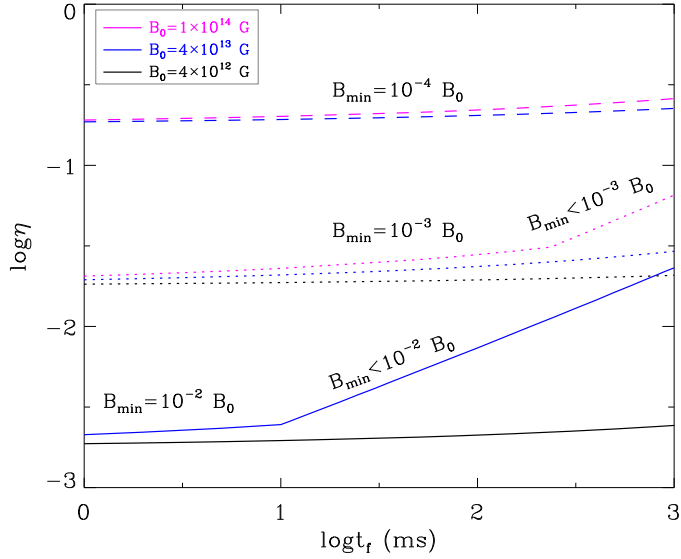}
  \caption{The required initial disk mass $\eta${\SM} as a function of the disk-formation timescale $t_{\rm f}$ for a given field decay, where $r_{\rm i}=2.5$, $r_{\rm f}=100$, and $B_{\min}$ is the minimal field-strength during the decay. The black, blue and magenta lines correspond to the initial fields $B_0=$ {4\TDP{12}}, {4\TDP{13}} and {\DP{14}} G, respectively. And the solid, dotted and dashed lines denote the case of $B_{\min}/B_{0}=$ {\DP{-2}}, {\DP{-3}} and {\DP{-4}}, respectively. Additionally, the weakest field decay for NSs with $B_{0}=$ {4\TDP{13}} G at $t_{\rm f} \gtrsim$ 10 ms ($B_{0}=$ {\DP{14}} G at $t_{\rm f} \gtrsim$ 240 ms) is plotted, i.e., $\max(B_{\min}/B_{0}) <$ {\DP{-2}} ($\max(B_{\min}/B_{0}) <$ {\DP{-3}}).}
  \label{fig06}
\end{center}
\end{figure}

\begin{figure}%[ht]
\begin{center}
  \includegraphics[width=0.8\textwidth]{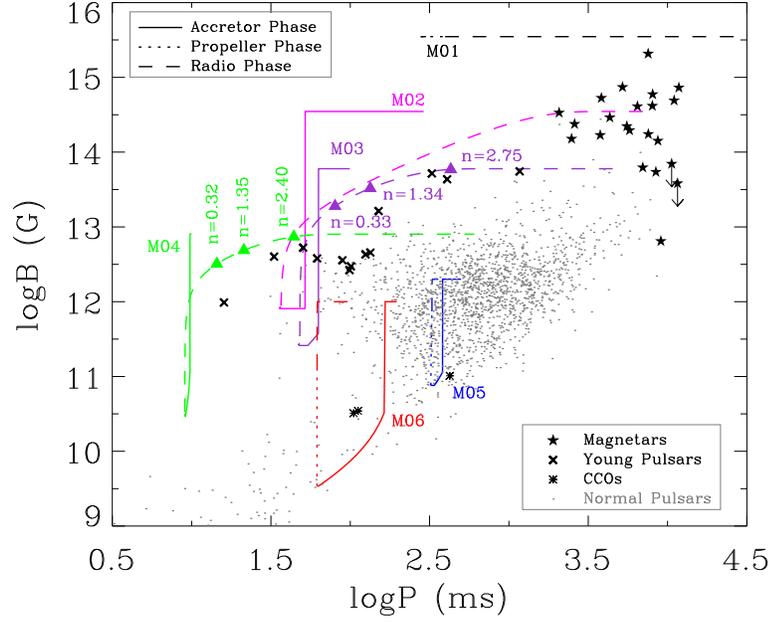}
  \caption{Six possible evolutionary tracks for young NSs in the $B-P$ diagram according to the fallback disk model, describing the possible evolution of CCOs, pulsars, magnetars, etc. Each evolutionary trajectory starts from the accreting phase (the solid line), and stops at $t=$ {\DP{5}} yr or {\DP{4}} yr for NSs with $B_{0} \lesssim$ {\DP{14}} G or $\gtrsim$ {\DP{14}} G. For models M03 and M04, several braking indices (ticked with filled triangles) in the ejector stage are calculated.}
  \label{fig07}
\end{center}
\end{figure}

\begin{enumerate}[1)]
  \item{In our model, it is the disk mass that mainly determines the occurrence and the extent of the field decay. And the effect of the field burial/reemergence may not be significant (i.e., the buried field is less than one percent of the initial field, see Fig.~\ref{fig06}), if the fallback disk is smaller than a few {\DP{-3}\SM}. In addition, with increasing initial magnetic field, the extent of field decay decreases, and the weakest decay becomes more and more significant (see Eq.~(\ref{eqn08})).}
  \item{According to the fallback disk model, NS sub-populations may evolve from NSs born with different disk mass ({\DP{-6}}--0.5 {\SM}) and magnetic field ($\gtrsim$ {\DP{12}} G), where the key point is the interaction between the fallback accretion and the magnetic field.}
  \item{If the NSs experience substantial field burial (i.e., initial disk mass $>$ a few {\DP{-3}\SM}), then during the reemergence phase the growth of the magnetic field leads to pulsars with the braking indices less than 3, which is mainly caused by the field diffusion, i.e., the second term on the right-hand side of Eq.~(\ref{eqn11}). Some NSs with initially high-$B$ field could evolve from normal pulsars to magnetars.}
  \item{NSs undergoing significant field burial have relatively weak fields, and it is difficult for them to recover their fields within a few {\DP{3}} yr. These NSs may appear as CCOs.}
  \item{If the NS is born with a disk $\lesssim$ {\DP{-4}\SM} and a field $\lesssim$ a few {\DP{15}} G, the fallback accretion can marginally affect the evolution of the spin or magnetic field. Thus the case may happen in some magnetars and most normal pulsars, since the magneto-dipolar radiation is the main braking torque.}
%  \item{}
%  \item{}
%  \item{}
\end{enumerate}

\begin{acknowledgements}
This work was supported by the National Key Research and Development Program of China (2016YFA0400803), the Natural Science Foundation of China under grant Nos. 11333004, 11773015, and 11573016, Project U1838201 supported by NSFC and CAS, and the Program for Innovative Research Team (in Science and Technology) at the University of Henan Province.
\end{acknowledgements}

\clearpage

\clearpage

\begin{table}%[H]
  \centering
  \caption{\label{tab01}Description of six representative models (in each model, $r_{\rm i}=2.5$, $r_{\rm f}=1000$) in Fig.~\ref{fig07}.}
  \vspace{0.2cm}
  \begin{footnotesize}%\small
  \begin{tabular}{lcccc}
  \hline
  Model &    $\eta$      & $t_{\rm f}$ & $P_{0}$ & $B_{0}$       \\
        &                &    (ms)     & (ms)    & ({\DP{14}} G) \\
  \hline
  M01   & 5.30{\TDP{-5}} &   5         &  300    &     35        \\
  M02   & 1.17{\TDP{-2}} &   2         &  300    &    3.5        \\
  M03   & 5.33{\TDP{-3}} &   5         &  100    &    0.6        \\
  M04   & 5.33{\TDP{-3}} &   5         &  10     &  {8\TDP{-2}}  \\
  M05   & 5.33{\TDP{-4}} &   5         &  500    &  {2\TDP{-2}}  \\
  M06   & 5.33{\TDP{-3}} &   5         &  200    &  {\DP{-2}}    \\
  \hline
  \end{tabular}
  \end{footnotesize}
\end{table}

\end{document}